# Improving Persian Document Classification Using Semantic Relations between Words

Saeed Parseh

Department of Software Engineering, Faculty of
Computer Engineering
University of Isfahan
Isfahan, Iran
saeedparseh@gmail.com

Ahmad Baraani

Department of Software Engineering, Faculty of
Computer Engineering
University of Isfahan
Isfahan, Iran
ahmadb@eng.ui.ac.ir

*Abstract*—With the increase of information, document classification as one of the methods of text mining, plays vital role in many management and organizing information. Document classification is the process of assigning a document to one or more predefined category labels. Document classification includes different parts such as text processing, term selection, term weighting and final classification. The accuracy of document classification is very important. Thus improvement in each part of classification should lead to better results and higher precision. Term weighting has a great impact on the accuracy of the classification. Most of the existing weighting methods exploit the statistical information of terms in documents and do not consider semantic relations between words. In this paper, an automated document classification system is presented that uses a novel term weighting method based on semantic relations between terms. To evaluate the proposed method, three standard Persian corpuses are used. Experiment results show 2 to 4 percent improvement in classification accuracy compared with the best previous designed system for Persian documents.

I. INTRODUCTION

The increase of information content has recently made it necessary to create a system for automatic classification of documents. Classification is one of the central issues in information systems dealing with text data. Document classification (DC) or Text classification (TC) is a supervised learning task of assigning natural language text documents to one or more predefined categories or classes according to their contents. The workflow in most DC systems is to train the classification system using a training dataset, including many text documents whose categories are known (training phase) then, assigning a category to a new document by this learned system (test phase).

The common paradigm of representing a document is the vector space model (VSM). Specifically, each document is transformed into a feature vector, where each feature refers to a term occurring in the document and the feature value corresponds to its weight. This weight represents what the term contributes to the semantics of the document d. Thus, there is an important issue in the text classification: how to weight a term. Different approaches have been introduced for term weighting [1-3]. These approaches vary in terms of the definition of a term and the determination of term weights. Most of these approaches exploit the statistical information of terms in documents and do not consider semantic relations between words. The usual bag-of-words (BOW) approach treats each word as a feature and considers the features independent of each other. Such an approach ignores the syntactic and semantic information in a document, such as word order, multi-word phrases, synonymy, polysemy, and other semantic relationships among words [4].

Most automatic text classification systems designed for English texts and they aren't useful for Persian documents. Development of Persian automated classification system due to the nature of Persian language is relatively difficult. In this paper, to improve accuracy of Persian text classification, a novel term weighting method is presented for Persian document, which weights terms by considering semantic relation of terms as a measure of dependency. Also, for determining the semantics of categories based on terms appearing in category labels, this method extend feature vector of each category by using a thesaurus.

The rest of this paper is organized as follows. Section II describes related works in Persian document classification. The process of document classification is described in section III. In section IV, the novel term weighting method is proposed. Implementation and results is demonstrated in section V. Finally, the paper is concluded in section VI.

II. RELATED WORKS

Improving accuracy of text classifiers has been an important issue and many studies have been conducted in this area. Much work has been conducted to find out effective approaches to represent document for text classification. Traditional "Bag of Words" (BOW) approach, which represents a document as a vector of weighted occurrence frequencies of individual terms, is limited because it only accounts for term frequency in the documents and can only use pieces of information that are explicitly mentioned in the training dataset. To overcome this limitation, some methods for extending the feature vector are developed. Most of these methods use an existing ontology or thesaurus [5]. Bloehdorn and Hotho represent documents using WordNet, the MeSH (Medical Subject Headings) Tree Structure Ontology [6]. They have shown that summarizing words in documents by synonyms in WordNet can improve the performance of TC. Hotho et al. utilized a term ontology structured from WordNet to improve the BOW text representation [7]. The authors adopted various strategies to enrich text document representation with synonyms and hyponyms from WordNet.

Because of the complex nature of Persian language, such as words with separate parts and combined verbs, the most of text classification systems are not applicable to Farsi texts. So previous works deal with to automated Farsi text classification is limited to next few works. Basiri et al. presented a comparison between KNN and fuzzy KNN approaches for Farsi text classification based on information gain and document frequency feature selection [8]. Pilevar et al.provided a Farsi text classification system using the Learning Vector Quantization network [9]. In this method, each class is presented by an essence vector called the codebook. These vectors are placed in the feature space in a manner that decision boundaries are approximated by the k-nearest neighbor (KNN) rule. Maghsoodi and Homayounpour have used SVM classifier based on extending the feature vector applying words extracted from a thesaurus [5]. This method has improved classifier performance when training dataset is unbalanced and not comprehensive for some classes. Elahimanesh et al. improved the KNN text classifier by inserting a factor to the KNN formula for considering the effects of unbalanced training datasets and used of N-grams with lengths more than 3 characters in text preprocessing [10]. Their approach improves the KNN algorithm especially when 8-grams indexing method and removing stop words are applied. Parchami et al. proposed a method that uses WordNet to increase similarity of documents under the same category [11]. Documents are represented by single words and their frequencies, by using WordNet, frequency of related words is changed to acquire higher accuracy.

### III. TEXT CLASSIFICATION

Text classification is defined as assigning predefined categories to text documents. In other words, the goal of this process is to find an outline from a document set D={$d_1$, $d_2$, ..., $d_i$} to a set of categories C ={$c_1$, $c_2$, . . . , $c_j$}. Commonly, four steps for construction of a text classifier can be considered:

- Document preprocessing and representation are the numeric representation of documents.
- Feature selection involves selecting some representative terms from all occurring terms in the documents to improve the efficiency of the classifier.
- Feature weighting assigns a numeric weight to each selected feature to more sufficiently represent document diversity.
- Classifier training: Numeric vectors are used to train the classifier.

Each of the above steps will discuss in this section.

#### A. Document Preprocessing

The preprocessing phase prepares the document for the classification process. Commonly the steps taken please for the document processing are: 1) Tokenization where a document is treated as a string, and then partitioned into a list of tokens. In fact, a document representation model is used. There are many different document representation models. The simplest is N-gram where words are represented as strings of length N. The most popular and effective representation is single words, where documents are represented by their words. 2) Removing stop words where words are frequently occurring and the insignificant words need to be removed. 3) Stemming word where a stemming algorithm that converts different word form into similar canonical form, is applied. This step is the process of conflating tokens to their root form.

In this paper, single words model for document representation is used but some special Persian language considerations must be regarded. The most important parameter is how to recognize words in the sentence. Automatic detection the words boundaries in Persian language is a very complex process. Persian language is one of the Hindi–European languages and the words usually are separated by space, but because of the existence of different scripts in this language, space isn't a definite and precise criterion for the diagnosis the word boundary. In Persian texts, there are two types of space: outside space and inside space.

- Outside space: spacing between words of a sentence or phrase that is considered a letter in computer.
- Inside space: Spacing between the morphemes consisting the word that called half-space.

In other words in Persian language, boundaries between the separated words is defined by a space and the boundaries of a set of morphemes that built a word are defined by a half-space; but this law necessarily not being fulfilled in all texts. Use of half-space, especially in manuscripts texts and the texts which typed with some text editor such as NotePad, WordPad and so on, is difficult and sometimes creates confusion in pronunciation. On the other hand the spacing rules are different in various scripts and according to different written formats of the letters in Persian language, all of various written forms of a word may be correct and usable in the texts. So creating a tokenizer that be able to recognize the words correctly is very important.

#### B. Term Selection

Each document in the training dataset consists of a great number of relevant and irrelevant terms corresponding to its category. So the second step in training phase is selecting the relevant terms. This step is called term selection (TS) or feature selection. The accuracy of the system depends highly on the keywords selected to represent documents, also the computation complexity depends on the number of keywords selected; choosing a small subset of a category relative keywords or choosing uninformative keywords that are not related to the domain of any involved category can lead to a deficiency in the accuracy of classification, also a very large number of keywords can make a classification algorithm inefficient in terms of time. In text classification, terms should be able to discriminate between categories and selected terms for different categories should not overlap. Some conventional term selection methods are information gain, $\chi^2$ test and document frequency [12-14].

In this paper, the mechanism of term selection is based on document frequency; that is, more repetition of a word in a category and less repetition of that word in other categories. In other words, the degree that a term corresponds to a specific category is defined as term frequency times inverse category frequency (*tf_icf*), according to (1) [5]:

$$tf\_icf(t_{ij}) = t_{ij} \times \log(\frac{\sum d_{ij}}{d_{ij}}) \quad (1)$$

Where $d_{ij}$ is the number of documents in which term i occurs in category j and $t_{ij}$ is the number of term i which occurs in category j. Therefore, a greater $t_{ij}$ indicates more repetition of a word in documents. Multiplying $t_{ij}$ by a category frequency reduces its value if the word appears in most of the categories. The next problem is finding an appropriate threshold for *tf_icf*. According to [5] and our study, $\frac{5}{\log_{10}|c|}$ is a good value for this threshold. In the threshold equation, |c| is the number of categories.

*C. Term Weighting*

The selected terms from the term-selection phase can be represented by the vector space model [15]. The elements of this vector are the term weights, which can be calculated using different weighting schemes. The most commonly used method is Term Frequency-Inverse Document Frequency (*tf_idf*), which stands for term frequency (*tf*) multiplied by inverse document frequency (*idf*), where *tf* shows the relative frequency of a certain term appearing in a document. The *tf_idf* weight for each term can obtain using (2) [16]. The weights resulting from *tf_idf* are often normalized by cosine normalization, given by (3) [16].

$$tf\_idf(t_i, d_j) = tf(t_i, d_j) \times \log(\frac{N}{N(t_i)}) \quad (2)$$

$$w_{ij} = \frac{tf\_idf(t_i, d_j)}{\sqrt{\sum_{s=1}^{|d_j|} tf\_idf(t_s, d_j)}} \quad (3)$$

Where N is the number of all documents, $|d_j|$ is the number of words in document $d_j$, $N(t_i)$ is the number of documents in the collection in which the term $t_i$ occurs at least once and $tf(t_i,d_j)$ is the frequency of the word $t_i$ in document $d_j$. Efficiency of this method on different corpuses has been proved [16]. In this paper, we use a novel term weighting method that is presented in the section IV.

*D. Classifier*

After obtaining feature vectors from the previous step, a machine learning method [16] or a statistical technique [17] is used as a classifier component. According to a study by [18], a support vector machine (SVM) outperforms other machine learning methods. It is based on the principle of structural risk minimization. In linear classification, SVM creates a hyper plane that separates the data into two sets with the maximum margin. A hyper plane with the maximum margin has the distances from the hyper plane to points when the two sides are equal. Mathematically, SVMs learn the sign function f (x) = sign(wx + b) , where w is a weighted vector in $R^n$. SVMs find the hyper plane y = wx + b by separating the space $R^n$ into two half spaces with the maximum margin. Linear SVMs can be generalized for non-linear problems. To do so, the data is mapped into another space H and we perform the linear SVM algorithm over this new space.

There is great interest to use SVM in classification systems. Therefore, in the approach presented in this paper, SVM is applied as implemented in the SVM_multicalss [19] software.

IV. PROPOSED METHOD

Before In this section, a novel term weighting method based on semantic relations between words and *tf-idf* is presented. The proposed method uses two parameters, semantic weight and *tf-idf* weight. Semantic weights are obtained with the help of a thesaurus. First, we present two definitions as follows:

**Definition 1.** (Similarity) If $S_i$ and $S_j$ are two sets of words, then similarity between them are obtained as follows:

$$Similarity(S_i, S_j) = Common\_Words(S_i, S_j) \quad (4)$$

Where Common_Words is a function computes the number of common words in two set of words.

For acquiring the semantic weight, for each term, all related terms to it, are extracted from thesaurus and put in a set what called "Semantic set" (SS). We use SSs for two goals: 1) Obtaining semantic weight. 2) Extending feature vectors to rich documents. In fact, a text may be expressed in different way using different words by human, so two texts are similar semantically, but are different lexically. Therefore, for resolving this problem, we rich documents using SSs.

**Definition 2.** (Semantic weight) If $t_i$ is a term in document $d_j$, then the semantic weight of $t_i$ in $d_j$ is:

$$Weight_{semantic}(t_i, d_j) = \sum_{i=1 \& w_i \neq w_j}^{n} Similarity(SS_{t_i}, SS_{t_j}) \quad (5)$$

Where n is the number of all terms in the document $d_j$ and is semantic set of term $t_i$ in $d_j$.

After acquiring all SSs for all terms in a document, similarity between these SSs are obtained using definition 1, then the semantic weight of each term is obtained using definition 2.

Then, using (2), *tf-idf* weight of each term is obtained and normalized by (3). Finally, we use (6) for term weighting.

$$Weight(t_i, d_j) = weight_{Semantic}(t_i, d_j) + tf\_idf(t_i, d_j) \quad (6)$$

After term weighting, there is a feature vector for each document in the data set. Then, as previously mentioned, we use semantic sets for extending feature vectors. If a feature is added to the feature vector from the feature selection (features that exist in data set), the weighting process is done according to (6), a feature is selected from the thesaurus, (7) [5] is used for term weighting.

$$Weight(w_{ij}) = \begin{cases} \frac{|d_{ij}|}{\sum_{j=1}^{|c|} |d_{ij}|} & ==> |d_{ij}| > 0 \\ \frac{1}{|c|} \end{cases} \quad (7)$$

Where $|d_{ij}|$ is the frequency of this feature in all the documents belonging to category j and $|c|$ is the number of categories.

In this paper, the "Farhang-e Teyfi" thesaurus [20] is used for acquiring the semantic weight and extending the feature vectors.

Finally, the classification system for Persian documents is designed in Fig. 1.

## V. IMPELEMENTATION AND EXPERIMENTAL RESULTS

A variety of experiment is conducted to test the performance and behavior of the proposed term weighting. In the following sections, the training and test data used in the experiments are explained, and then obtained results are presented and analyzed.

### A. Training Data Set

In our experiments, we have used three corpuses: "Hamshahri" (version 2) [21], "Persica" [22] and "Tabnak". The Hamshahri corpus is undoubtedly the largest Persian text collection containing more than 310,000 documents with the following subject categories: politics, urban news, economics, reports, editorials, literature, sciences, society, foreign news, sports, etc., from 1996 to 2007. Persica and Tabnak, are other corpuses with the same categories as Hamshahri.

To evaluate of the proposed method, three different data sets in six categories are selected from above corpuses randomly. The first data set (D1) with 585 random documents, the second (D2) with 976 random documents and the third (D3) with 1225 random documents are selected. Table 1 represents the name and number of documents used in these data sets. The purpose of using three different data sets is to provide reasonable results.

### B. Evaluation Measures

Precision (8) and recall (9), measures are widely used for evaluation of the classification tasks. They are defined as follows [16]:

$$\Pr ecision = \frac{TP}{TP + FP} \quad (8)$$

$$\operatorname{Re} call = \frac{TP}{TP + FN} \quad (9)$$

Where TP is the number of documents correctly assigned to a category, FP is the number of documents incorrectly assigned to a category and FN is the number of documents incorrectly omitted from a category. There is a trade-off between precision and recall of a system [16]. The F1-measure, (10), is the harmonic mean of precision and recall that takes into account effects of both precision and recall measures.

$$F1\_measure(r, p) = \frac{2 \times r \times p}{r + p} \quad (10)$$

In this paper, average precision and recall measures are used and their equations are as follows [16]:

$$p_{ave} = \frac{\sum_{i=1}^{|c|} p(c_i)}{|c|} \quad (11)$$

$$r_{ave} = \frac{\sum_{i=1}^{|c|} r(c_i)}{|c|} \quad (12)$$

$$F1\_measure_{ave} = \frac{2 \times r_{ave} \times p_{ave}}{r_{ave} + p_{ave}} \quad (13)$$

Where, $p(c_i)$ and $r(c_i)$ are the precision and recall of category $i$.

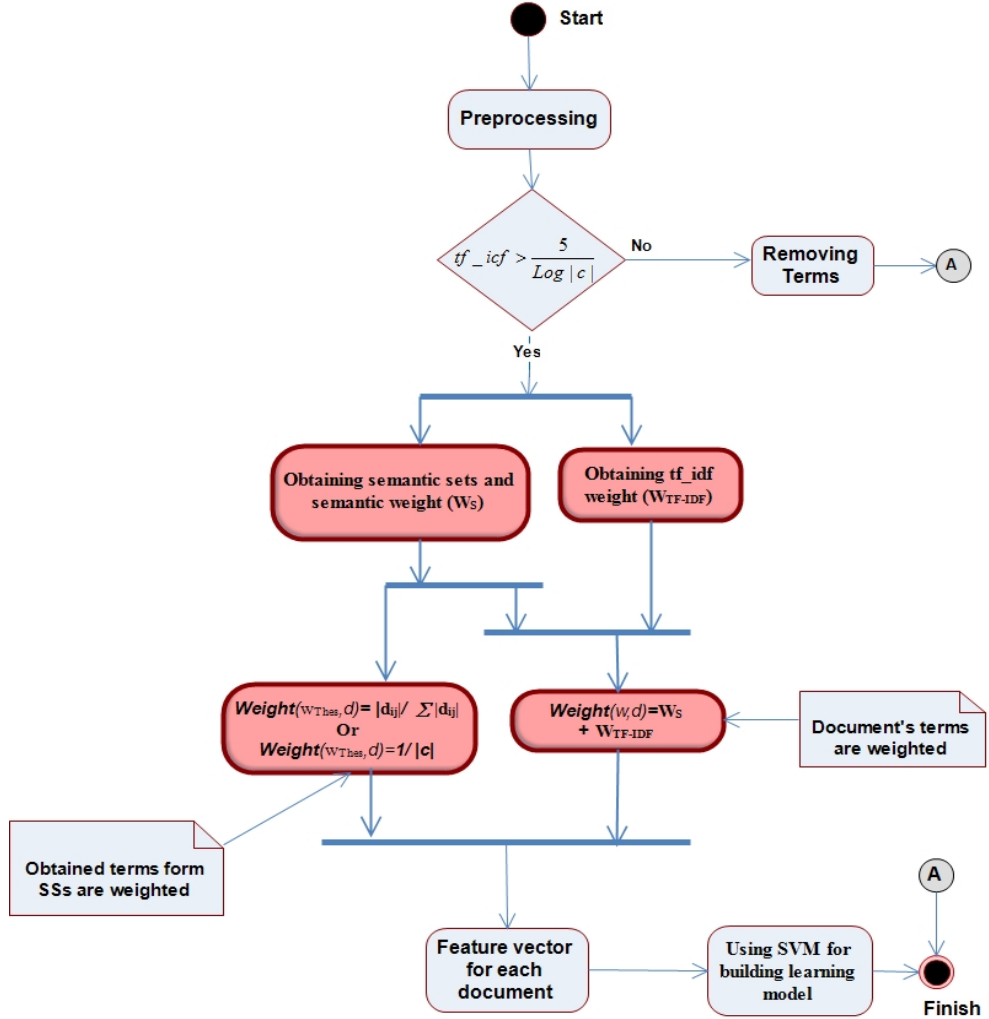

Figure 1.  Activity diagram of designed document classification

TABLE 1.  DETAILED DESCRIPTION OF THREE USED DATA SETS

| Class Name | D1 | | D2 | | D3 | |
|---|---|---|---|---|---|---|
| | *No. of training docs* | *No. of testing docs* | *No. of training docs* | *No. of testing docs* | *No. of training docs* | *No. of testing docs* |
| Society | 64 | 51 | 105 | 60 | 167 | 57 |
| Economy | 63 | 54 | 109 | 63 | 175 | 64 |
| Politic | 71 | 49 | 103 | 62 | 159 | 50 |
| Culture | 64 | 54 | 114 | 63 | 174 | 54 |
| Medicine | 56 | 35 | 94 | 44 | 159 | 57 |
| Sport | 55 | 49 | 100 | 59 | 151 | 65 |
| **Total** | **373** | **212** | **625** | **351** | **985** | **240** |

## C. Equations

In this subsection, the experimental results are presented. The experiments consist of evaluating classifier performance when thesaurus and semantic weight are used, using three data sets mentioned in table 1.

As shown in table 2, with extending feature vectors, the classification performance improves especially when accessible training datasets are not sufficiently comprehensive; i.e. the existing words are not able to distinguish a category from other categories. In this case, the use of a subsidiary knowledge resource such as a thesaurus may compensate the insufficiency in the existing information. In fact, some relevant features (words) obtained from the thesaurus is added to the existing feature vector which has been obtained by processing the training documents belonging to the desired category. When the semantic weight is used in process of term weighting (the proposed method), efficiency measurements increase about 2 to 4 percent.

TABLE 2. COMPARISON OF USING AND NOT USING THESAURUS AND PROPOSED METHOD

| Dataset | Using tf-idf without semantic weight and extending feature vectors | | | Using tf-idf without semantic weight, with extending feature vectors | | | Proposed method | | |
|---|---|---|---|---|---|---|---|---|---|
| | Precision | Recall | F1-measure | Precision | Recall | F1-measure | Precision | Recall | F1-measure |
| **D1** | 73.62 | 69.26 | 71.37 | 97.28 | 97.05 | 97.17 | **99.07** | **99.07** | **99.07** |
| **D2** | 79.52 | 76.54 | 78 | 91.52 | 87.67 | 89.55 | **93.82** | **92.47** | **93.14** |
| **D3** | 80.15 | 77.12 | 78.61 | 92.67 | 89.12 | 90.86 | **94.04** | **91.75** | **92.88** |

## D. Comparison With Other Works

In this subsection, Persian document classification system in this paper is compared with previous systems designed for Persian documents. For this purpose, we selected three previous works that are similar to the system in this paper and have reported the best result. These works mentioned in section II. We refer to them according to the following

1) System 1: The TC system designed by Maghsoodi and Homayounpour [5] using SVM and thesaurus.
2) System 2: The TC system proposed by Parchami et al. [11] using SVM and thesaurus.
3) System 3: The TC system designed by The Elahimanesh et al. [10] using k-NN.

After implementing them and test on three data sets, the results are shown in figure 2. Since F1-measure is harmonic mean of precision and recall, so this measure has selected for comparison.

As shown in figure 2, the designed document classification system in this paper has provided the better result relative to other systems according to F-measure. This is because, the designed system in this paper use semantic relations between words for weighting them.

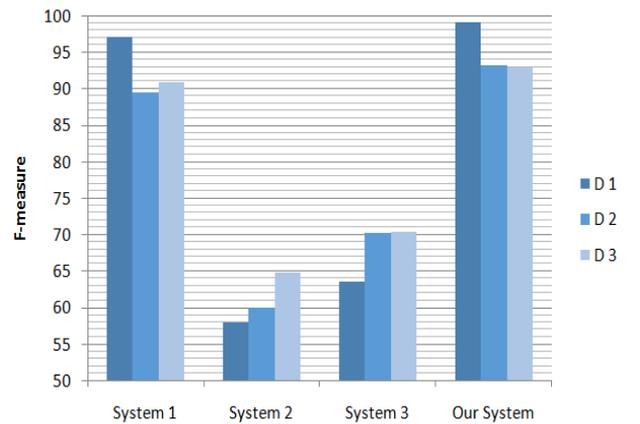

Figure 2. Comparison of other document classification systems with our system

The semantic relations are specified according to common meaning between words. The number of shared meanings of a word with other words is considered as semantic weight. More repetition of a word in a document and less repetition of that word in other documents shows importance of that word for the document. So $tf\_idf$ weight is used for considering this importance. Using semantic and statistical information for weighting as shown in Fig. 2, improve the accuracy of classification. System 1 only uses thesaurus to rich document and $tf\_idf$ for weighting and does not consider other information such as semantic relations. Also, system 2 and 3 use statistical information for weighting, so their accuracy is lower than the proposed system in this paper.

## VI. CONLUSION

In this paper, we have proposed a new term weighting method to improve accuracy of Persian text classification. The proposed method acquires semantic relations between terms using a thesaurus to obtain a semantic weight for each term in a document and using *tf-idf* method, obtains a statistical weight for each term, finally uses the sum of these weights for weighting terms. In this method, to rich documents, feature vector of each document is extended using thesaurus. The obtained results indicated that the proposed method is able to increase the performance of average recall, precision and F1 criteria.

## REFERENCES

[1] C. Buckley, A. Singhal, and M. Mitra, "New retrieval approaches using SMART," In Proc. of the 4th Text Retrieval conference (TREC-4), Gaithersburg, 1996, pp. 25-48.

[2] K. S. Jones and P. Willett, Readings in information retrieval, San Francisco: Morgan Kaufmann Publishers, 1997, pp. 85-167.

[3] G. Salton and C. Buckley, "Term-weighting approaches in automatic text retrieval," Journal of Information Processing and management, vol. 24, 1988, pp. 513-523.

[4] O. Luo, CH. Enhong, and X. Hui, "A Semantic Term Weighting Scheme for Text Categorization," Expert Systems with Applications, vol. 38, 2011, pp. 12708-12716.

[5] N. Maghsoodi and M. Homayounpoor, "Using Thesaurus to Improve Multiclass Text Classification," In Computational Linguistics and Intelligent Text Processing, Springer Berlin Heidelberg, 2011, pp. 244-253.

[6] S. Bloehdorn and A. Hotho, "Boosting for text classification with semantic features," In Workshop on Text-based Information Retrieval (TIR 2004) at the 27th German Conference on Artificial Intelligence, 2004, pp. 149–166.

[7] Hotho, S. Staab, and G. Stumme, "Wordnet improves text document clustering," In Proceedings of the Semantic Web Workshop at SIGIR, 2003, pp. 61–69.

[8] M. E. Basiri, S. Nemati, and N. Aqaee, "Comparing KNN and FKNN algorithms in Farsi text classification based on information gain and document frequency feature selection," In Proceedings of the 13th International Computer Conference of Computer Society of Iran, 2008, pp. 383-406.

[9] M. T. Pilevar, H. Feili, and M. Soltani, "Classification of Persian textual documents using learning vector quantization," In Proceedings of the International Conference on Natural Language Processing and Knowledge Engineering, 2009, pp. 1-6.

[10] M. H. Elahimanesh, B. Minaei, and H. Malekinezhad, "Improving K-Nearest Neighbor Efficacy for Farsi Text Classification," The International Conference on Language Resources and Evaluation (LREC), 2012, pp. 1618-1621.

[11] M. Parchami, B. Akhtar, and M. Dezfoulian, "Persian text classification based on K-NN using wordnet," In Advanced Research in Applied Artificial Intelligence, Springer Berlin Heidelberg, 2012, pp. 283-291.

[12] Y. L. Huang, "A theoretic and empirical research of cluster indexing for Mandarin Chinese full text document," Journal of Library and Information Science, vol. 24, 1998, pp. 1023–2125.

[13] C. Lee and G. Lee, "Information gain and divergence-based feature selection for machine learning-based text categorization," Information Processing & Management, vol. 42, 2006, pp. 155–165.

[14] Y. Yang and I. O. Pedersen, "A comparative study on feature selection in text categorization," In Proceedings of 14th International Conference on Machine Learning (ICML-97), San Francisco,1997, pp. 412–420.

[15] G. Salton, C. Yang, and A, Wang, "A vector space model for automatic indexing," Communications of the ACM, vol. 18, 1975, pp. 613–620.

[16] F. Sebastiani, "Machine learning automated text categorization," ACM Computing Surveys, vol. 34, 2002, pp. 1–47.

[17] Y. Yang, "Expert network: Effective and efficient learning from human decisions in text categorization and retrieval," In Proceedings of the International Conference on Research and Development in Information Retrieval (ACM SIGIR '94), NewYork, ACM Press, 1994, pp. 13–22.

[18] Y. Yang,"An evaluation of statistical approaches to text categorization," Information Retrieval, 1999, pp. 69–90.

[19] Th. Joachims. "Multi-Class Support Vector Machine. Internet:www.cs.cornell.edu/people/tj/svm_light/svm_multiclass.html, Aug. 14, 2008 [Oct. 1, 2014].

[20] J. Fararoy, Farhang-e Teyfi, Tehran: Hermes Press, 2008.

[21] AleAhmad, H. Amiri, E. Darrudi, M. Rahgozar, and F. Oroumchian, "Hamshahri: A Standard Persian Text Collection," Knowledge-Based Systems, vol. 22, 2009, pp. 382–387.

[22] H. Eghbalzadeh, B. Hosseini, Sh. Khadivi, and A. Khodabakhsh, "Persica: A Persian corpus for multipurpose text mining and Natural language processing," Sixth International Symposium on Telecommunications (IST), 2012, pp. 1207-1214.